\documentclass[preprint,prd,aps,showpacs,showkeys,nofootinbib]{revtex4}
\usepackage{graphicx}
\begin{document}

%\preprint{hep-ph/0412147}

\title{Top quark decays with flavor violation in the B-LSSM}

\author{Jin-Lei Yang$^{a}$\footnote{JLYangJL@163.com},
Tai-Fu Feng$^{a}$\footnote{fengtf@hbu.edu.cn},
Hai-Bin Zhang$^{a}$\footnote{hbzhang@hbu.edu.cn},
Guo-Zhu Ning$^{a}$\footnote{ninggz@hbu.edu.cn},
Xiu-Yi Yang$^{b}$}

\affiliation{$^a$Department of Physics, Hebei University, Baoding, 071002, China\\
$^b$Department of Science, University of Science and Technology Liaoning, Anshan, 114051, China}

\begin{abstract}
The decays of top quark $t\rightarrow c\gamma,\;t\rightarrow cg,\;t\rightarrow cZ,\;t\rightarrow ch$ are extremely rare processes in the standard model (SM). The predictions on the corresponding branching ratios in the SM are too small to be detected in the future, hence any measurable signal for the processes at the LHC is a smoking gun for new physics. In the extension of minimal supersymmetric standard model with an additional local $U(1)_{B-L}$ gauge symmetry (B-LSSM), new gauge interaction and new flavor changing interaction affect the theoretical evaluations on corresponding branching ratios of those processes. In this work, we analyze those processes in the B-LSSM, under a minimal flavor violating assumption for the soft breaking terms. Considering the constraints from updated experimental data, the numerical results imply $Br(t\rightarrow c\gamma)\sim5\times10^{-7}$, $Br(t\rightarrow cg)\sim2\times10^{-6}$, $Br(t\rightarrow cZ)\sim4\times10^{-7}$ and $Br(t\rightarrow ch)\sim3\times10^{-9}$ in our chosen parameter space. Simultaneously, new gauge coupling constants $g_{_B},\;g_{_{YB}}$ in the B-LSSM can also affect the numerical results of $Br(t\rightarrow c\gamma,\;cg,\;cZ,\;ch)$.
\end{abstract}


\maketitle

\section{Introduction\label{sec1}}
\indent\indent
Since the running LHC provides an opportunity to seek out top quark rare decays,
top quark shows great promise in revealing the secret of new physics beyond the
standard model (SM). The branching ratios of the flavor changing neutral
current (FCNC) of top quark $t\rightarrow c\gamma$, $t\rightarrow cg$,
$t\rightarrow ch$, $t\rightarrow cZ$ in the SM are highly suppressed~\cite{SM1,SM2,SM3,SM4,SM5,SM6}
\begin{eqnarray}
&&{\rm Br}(t\rightarrow c\gamma)\sim5\times10^{-14},\quad\; {\rm Br}(t\rightarrow cg)\sim5\times10^{-12}, \nonumber\\
&&{\rm Br}(t\rightarrow ch)\sim3\times10^{-15},\quad\; {\rm Br}(t\rightarrow cZ)\sim1\times10^{-14},
\end{eqnarray}
and beyond the detection capabilities of LHC in the near future. Nevertheless the exotic mechanism
from the new physics can enhance those branching ratios drastically~\cite{FCNC}, which can
be detected in the future. The updated upper limits on the branching ratios of LHC are \cite{a1,a2,a3,a4,a5,a6,a7,a8,a9}
\begin{eqnarray}
&&{\rm Br}(t\rightarrow c\gamma)<1.7\times10^{-3},\quad\; {\rm Br}(t\rightarrow cg)<2\times10^{-4}, \nonumber\\
&&{\rm Br}(t\rightarrow ch)<2.2\times10^{-3},\quad\; {\rm Br}(t\rightarrow cZ)<2.3\times10^{-4}.
\end{eqnarray}

Hence, detecting those rare top quark decays on
the LHC provides a good window to search the new physics beyond the SM. Actually
several extensions of the SM predict the branching ratios of the rare top decays
surpassing the SM evaluations several orders numerically.
In Table~\ref{tab1}, we present the theoretical predictions on the branching ratios
of those rare decays of top quark in some popular new physics extensions of the SM.
Those new physics models include the two-Higgs doublet models
with flavour-conservation (FC 2HDM)~\cite{THDMFC1,THDMFC2,THDMFC3,THDMFC4} and without flavour-conservation (NFC 2HDM)~\cite{THDMNFC1}, minimal supersymmetric extension of the SM (MSSM)~\cite{MSSM1,Guasch:1999jp,MSSM2}, supersymmetry (SUSY) without R-parity~\cite{SUSY1,SUSY2},
the Topcolour-assisted Technicolour model (TC2)~\cite{3} and the extension with warped extra dimensions (RS)~\cite{4}. The results in Refs.\cite{MSSM1,Guasch:1999jp,MSSM2} were obtained when the supersymmetric particles were not constrained strongly by direct searches at the LHC. In addition, the results in Refs.~\cite{MSSM1,Guasch:1999jp} under a minimal flavor violating assumption for the soft breaking terms, while Ref.\cite{MSSM2} takes into account the off-diagonal terms for the squark matrices. Hence the results in Ref.\cite{MSSM2} are larger than those in Refs.~\cite{MSSM1,Guasch:1999jp}.
\begin{table*}
\begin{tabular*}{\textwidth}{@{\extracolsep{\fill}}lllllllll@{}}
\hline
Process & 2HDM\cite{THDMFC1,THDMFC2,THDMFC3,THDMFC4} & FC 2HDM\cite{THDMNFC1} & MSSM\cite{MSSM1,Guasch:1999jp} & MSSM\cite{MSSM2} & $\not{R}$ SUSY\cite{SUSY1,SUSY2} & TC2\cite{3} & RS\cite{4}\\
\hline
$t\rightarrow q\gamma$ & $\sim10^{-6}$ & $\sim10^{-9}$ & $\sim10^{-8}$ & $\sim10^{-6}$ & $\sim10^{-6}$ & $\sim10^{-6}$ & $\sim10^{-9}$\\
$t\rightarrow qg$ & $\sim10^{-4}$ & $\sim10^{-8}$ & $\sim10^{-6}$ & $\sim10^{-5}$ & $\sim10^{-4}$ & $\sim10^{-4}$ & $\sim10^{-9}$\\
$t\rightarrow qh$ & $\sim10^{-3}$ & $\sim10^{-5}$ & $\sim10^{-8}$ & $\sim10^{-5}$ & $\sim10^{-6}$ & ----- & -----\\
$t\rightarrow qZ$ & $\sim10^{-7}$ & $\sim10^{-10}$ & $\sim10^{-8}$ & $\sim10^{-6}$ & $\sim10^{-7}$ & $\sim10^{-4}$ & $\sim10^{-5}$\\
\hline
\end{tabular*}
\caption{FCNC decays branching ratios of several SM extension.}
\label{tab1}
\end{table*}

In the supersymmetric extensions of the SM, the extension with local $B-L$ gauge symmetry
(here employing the abbreviation B-LSSM denoting the extension)
draw the attention of physicists, since it provides the candidate for the cold dark matter
and the seesaw mechanism to explain the oscillation of neutrinos naturally.
The model B-LSSM~\cite{5,6} is based on the gauge symmetry group $SU(3)_C\otimes SU(2)_L\otimes
U(1)_Y\otimes U(1)_{B-L}$, where $B$ stands for the baryon number and
$L$ stands for the lepton number respectively.
Besides accounting elegantly for the existence and smallness of the left-handed neutrino masses,
the B-LSSM also alleviates the aforementioned little hierarchy problem of the MSSM~\cite{search}, because the exotic singlet Higgs and
right-handed (s)neutrinos~\cite{7,77,88,99,10,11} release additional
parameter space from the LEP, Tevatron and LHC constraints.

In this model, the invariance under $U(1)_{B-L}$ gauge group imposes the the R-parity conservation which is assumed in the MSSM to avoid proton decay. And R-parity conservation can be maintained if $U(1)_{B-L}$ symmetry is broken spontaneously~\cite{C.S.A}. Furthermore, it could help to understand the origin of R-parity and its possible spontaneous violation in the supersymmetric models~\cite{S.K,P.F,V.B} as well as the mechanism of leptogenesis~\cite{J.P,K.S.B}. Moreover, the model can provide much more candidates for the Dark Matter comparing that with the MSSM~\cite{16,1616,DelleRose:2017ukx,DelleRose:2017uas}.

In this work, we analyze those processes in the B-LSSM, under a minimal flavor violating assumption for the soft breaking terms. In this case, the only source of flavor violation comes from the Cabibbbo-Kobayashi-Maskawa (CKM) matrix in the quark sector. And we can explore the effects of new parameters to those processes, with respect to the MSSM. Our presentation is organized as follows. In Sec.~\ref{sec2}, the main ingredients of B-LSSM are summarized briefly by introducing the superpotential, the general soft breaking terms and the Higgs sector. In Sec.~\ref{sec3}, the branching ratios for $t\rightarrow c\gamma, t\rightarrow cg, t\rightarrow ch$ and $t\rightarrow cZ$ is calculated in the model. The numerical analyses are given in Sec.~\ref{sec4}, and Sec.~\ref{sec5} gives a summary.

\section{The B-LSSM\label{sec2}}
\indent\indent
In the B-LSSM, one enlarges the local gauge group of the SM to $SU(3)_C\otimes SU(2)_L\otimes U(1)_Y\otimes U(1)_{B-L}$, where the $U(1)_{B-L}$ is spontaneously broken by the chiral singlet superfields $\hat{\eta}_1$ and $\hat{\eta}_2$. In literatures there are several popular versions of B-LSSM. Here we adopt the version described in Refs.~\cite{44,Abdallah:2014fra,8,Khalil:2015wua,Hammad:2016trm} to proceed our analysis, while this version of B-LSSM is encoded in SARAH~\cite{164,165,166,167,168} which is used to create the mass matrices and interaction vertexes in the B-LSSM. Besides the superfields of the MSSM, the exotic superfields of the B-LSSM are three generations right-handed neutrinos $\hat{\nu}_i^c\sim$(1, 1, 0, 1) and two chiral singlet superfields $\hat{\eta}_{1}\sim(1,1,0,-1)$, $\hat{\eta}_{2}\sim(1,1,0,1)$. Meanwhile, quantum numbers of the matter chiral superfields for quarks and leptons are given by
\begin{eqnarray}
&&\hat{Q}_i\sim(3, 2, 1/6, 1/6), \quad\;\hat{L}_i\sim(1, 2, -1/2, -1/2), \quad\;\hat{U}_i\sim(3, 1, -2/3, -1/6),\nonumber\\
&& \qquad\; \quad\; \quad\;\hat{D}_i\sim(3, 1, 1/3, -1/6), \quad\;\hat{E}_i\sim(1, 1, 1, 1/2)
\end{eqnarray}
with $i=1,2,3$ denoting the index of generation. In addition, the quantum numbers of two Higgs doublets is assigned as
\begin{eqnarray}
&&\hat{H_1}=\left(\begin{array}{c}H_1^1\\ H_1^2\end{array}\right)\sim (1, 2, -1/2, 0),\quad\;\hat{H_2}=\left(\begin{array}{c}H_2^1\\ H_2^2\end{array}\right)\sim(1, 2, 1/2, 0).
\end{eqnarray}

The corresponding superpotential of the B-LSSM is written as
\begin{eqnarray}
&&W=W_{MSSM}+W_{(B-L)}.
\end{eqnarray}
Here, $W_{MSSM}$ is the superpotential of the MSSM, and $W_{(B-L)}$ is the sector involving exotic superfields, and
\begin{eqnarray}
&&W_{(B-L)}=Y_{\nu, ij}\hat{L_i}\hat{H_2}\hat{\nu}^c_j-\mu' \hat{\eta}_1 \hat{\eta}_2
+Y_{x, ij} \hat{\nu}_i^c \hat{\eta}_1 \hat{\nu}_j^c,
\end{eqnarray}
where $i, j$ are generation indices. Correspondingly, the soft breaking terms of the B-LSSM are generally given as
\begin{eqnarray}
&&\mathcal{L}_{soft}=\mathcal{L}_{MSSM}+\Big[-M_{BB^{'}}\tilde{\lambda}_{B^{'}} \tilde{\lambda}_{B} -
\frac{1}{2}M_{B^{'}}\tilde{\lambda}_{B^{'}} \tilde{\lambda}_{B^{'}} -B_{\mu^{'}}\tilde{\eta}_1 \tilde{\eta}_2 +T_{\nu}^{ij} H_2 \tilde{\nu}_i^c \tilde{L}_j+T_x^{ij} \tilde{\eta}_1 \tilde{\nu}_i^c \tilde{\nu}_j^c
\nonumber\\
&&\hspace{1.4cm}
+h.c.\Big]-m_{\tilde{\eta}_1}^2 |\tilde{\eta}_1|^2-m_{\tilde{\eta}_2}^2 |\tilde{\eta}_2|^2-m_{\tilde{\nu},ij}^2(\tilde{\nu}_i^c)^* \tilde{\nu}_j^c,
\end{eqnarray}
with $\lambda_{B}, \lambda_{B^{'}}$ denoting the gaugino of $U(1)_Y$ and $U(1)_{(B-L)}$ respectively,
$\mathcal{L}_{MSSM}$ is the soft breaking terms in MSSM. The local gauge symmetry $SU(2)_L\otimes U(1)_Y\otimes U(1)_{B-L}$ breaks down to the electromagnetic symmetry $U(1)_{em}$ as the Higgs fields receive vacuum expectation values (VEVs):
\begin{eqnarray}
&&H_1^1=\frac{1}{\sqrt2}(v_1+{\rm Re}H_1^1+i{\rm Im}H_1^1),
\qquad\; H_2^2=\frac{1}{\sqrt2}(v_2+{\rm Re}H_2^2+i{\rm Im}H_2^2),\nonumber\\
&&\tilde{\eta}_1=\frac{1}{\sqrt2}(u_1+{\rm Re}\tilde{\eta}_1+i{\rm Im}\tilde{\eta}_1),
\qquad\;\quad\;\tilde{\eta}_2=\frac{1}{\sqrt2}(u_2+i{\rm Re}\tilde{\eta}_2+i{\rm Im}\tilde{\eta}_2)\;.
\end{eqnarray}
For convenience, we define $u^2=u_1^2+u_2^2,\; v^2=v_1^2+v_2^2$ and $\tan\beta^{'}=\frac{u_2}{u_1}$ in analogy to the ratio of the MSSM VEVs ($\tan\beta=\frac{v_2}{v_1}$).

The presence of two Abelian groups gives rise to a new effect absent in the MSSM or other SUSY models with just one Abelian gauge group: the gauge kinetic mixing. It results from the invariance principle allows the Lagrangian to include a mixing term between the strength tensors of gauge fields associated with the $U(1)$ gauge groups, $-\kappa_{_{Y,BL}}A_{_\mu}^{\prime Y}A^{\prime\mu, BL}$, where $A_{_\mu}^{\prime Y}, A^{\prime\mu, BL}$ denote the gauge fields associated with the two $U(1)$ gauge groups, $Y, B-L$ corresponding to the hypercharge and B-L charge respectively, $\kappa_{_{Y,BL}}$ is an antisymmetric tensor which includes the mixing of $U(1)_Y$ and $U(1)_{B-L}$ gauge fields. This mixing couples the B-L sector to the MSSM sector, and even if it is set to zero at $M_{GUT}$, it can be induced through RGEs\cite{RGE1,RGE2,RGE3,RGE4,RGE5,RGE6,RGE7}.  In practice, it turns out that it is easier to work with non-canonical covariant derivatives instead of off-diagonal field-strength tensors. However, both approaches are equivalent\cite{R.F}. Hence in the following, we consider covariant derivatives of the form
\begin{eqnarray}
%%%%%%%%%%%%%%%%%%%%%%%%%%%%%%%%%%%%%%%%%%%%%%%%%%%%%%%%%%%%%%%%%%%%
&&D_\mu=\partial_\mu-i\left(\begin{array}{cc}Y,&B-L\end{array}\right)
\left(\begin{array}{cc}g_{_Y},&g_{_{YB}}^{'}\\g_{_{BY}}^{'},&g_{_{B-L}}\end{array}\right)
\left(\begin{array}{c}A_{_\mu}^{\prime Y} \\ A_{_\mu}^{\prime BL}\end{array}\right)\;.
%%%%%%%%%%%%%%%%%%%%%%%%%%%%%%%%%%%%%%%%%%%%%%%%%%%%%%%%%%%%%%%%%%%%
\label{gauge1}
\end{eqnarray}
As long as the two Abelian gauge groups are unbroken, we still have the freedom to perform a change of the basis
\begin{eqnarray}
%%%%%%%%%%%%%%%%%%%%%%%%%%%%%%%%%%%%%%%%%%%%%%%%%%%%%%%%%%%%%%%%%%%%
&&D_\mu=\partial_\mu-i\left(\begin{array}{cc}Y,&B-L\end{array}\right)
\left(\begin{array}{cc}g_{_Y},&g_{_{YB}}^{'}\\g_{_{BY}}^{'},&g_{_{B-L}}\end{array}\right)R^TR
\left(\begin{array}{c}A_{_\mu}^{\prime Y} \\ A_{_\mu}^{\prime BL}\end{array}\right)\;,
%%%%%%%%%%%%%%%%%%%%%%%%%%%%%%%%%%%%%%%%%%%%%%%%%%%%%%%%%%%%%%%%%%%%
\label{gauge2}
\end{eqnarray}
where $R$ is a $2\times2$ orthogonal matrix. Choosing $R$ in a proper form, one can write the coupling matrix as
\begin{eqnarray}
%%%%%%%%%%%%%%%%%%%%%%%%%%%%%%%%%%%%%%%%%%%%%%%%%%%%%%%%%%%%%%%%%%%%
&&\left(\begin{array}{cc}g_{_Y},&g_{_{YB}}^{'}\\g_{_{BY}}^{'},&g_{_{B-L}}\end{array}\right)
R^T=\left(\begin{array}{cc}g_{_1},&g_{_{YB}}\\0,&g_{_{B}}\end{array}\right)\;,
%%%%%%%%%%%%%%%%%%%%%%%%%%%%%%%%%%%%%%%%%%%%%%%%%%%%%%%%%%%%%%%%%%%%
\label{gauge3}
\end{eqnarray}
where $g_{_{1}}$ corresponds to the measured hypercharge coupling which is modified in
B-LSSM as given along with $g_{_{B}}$ and $g_{_{YB}}$ in~\cite{BLSSM1}. Then, we can redefine the $U(1)$ gauge fields
\begin{eqnarray}
%%%%%%%%%%%%%%%%%%%%%%%%%%%%%%%%%%%%%%%%%%%%%%%%%%%%%%%%%%%%%%%%%%%%
&&R\left(\begin{array}{c}A_{_\mu}^{\prime Y} \\ A_{_\mu}^{\prime BL}\end{array}\right)
=\left(\begin{array}{c}A_{_\mu}^{Y} \\ A_{_\mu}^{BL}\end{array}\right)\;.
%%%%%%%%%%%%%%%%%%%%%%%%%%%%%%%%%%%%%%%%%%%%%%%%%%%%%%%%%%%%%%%%%%%%
\label{gauge4}
\end{eqnarray}

Immediate interesting consequence of the gauge kinetic mixing arise in  various sectors of the model as discussed in the subsequent analysis. Firstly, $A^{BL}$ boson mixes at the tree level with the $A^Y$ and $V^3$ bosons. In the basis $(A^Y, V^3, A^{BL})$, the corresponding mass matrix reads,
\begin{eqnarray}
&&\left(\begin{array}{*{20}{c}}
\frac{1}{8}g_{_1}^2 v^2 & -\frac{1}{8}g_{_1}g_{_2} v^2 & \frac{1}{8}g_{_1}g_{_{YB}} v^2 \\ [6pt]
-\frac{1}{8}g_{_1}g_{_2} v^2 & \frac{1}{8}g_{_2}^2 v^2 & -\frac{1}{8}g_{_2}g_{_{YB}} v^2\\ [6pt]
\frac{1}{8}g_{_1}g_{_{YB}} v^2 & -\frac{1}{8}g_{_2}g_{_{YB}} v^2 & \frac{1}{8}g_{_{YB}}^2 v^2+\frac{1}{8}g_{_{B}}^2 u^2
\end{array}\right).\label{gauge matrix}
\end{eqnarray}
This mass matrix can be diagonalized by a unitary mixing matrix, which can be expressed by two mixing angles $\theta_{_W}$ and $\theta_{_W}'$ as
\begin{eqnarray}
&&\left(\begin{array}{*{20}{c}}
\gamma\\ [6pt]
Z\\ [6pt]
Z'
\end{array}\right)=
\left(\begin{array}{*{20}{c}}
\cos\theta_{_W} & \sin\theta_{_W} & 0 \\ [6pt]
-\sin\theta_{_W}\cos\theta_{_W}' & \cos\theta_{_W}\cos\theta_{_W}' & \sin\theta_{_W}'\\ [6pt]
\sin\theta_{_W}\sin\theta_{_W}' & -\cos\theta_{_W}'\sin\theta_{_W}' & \cos\theta_{_W}'
\end{array}\right)
\left(\begin{array}{*{20}{c}}
A^Y\\ [6pt]
V^3\\ [6pt]
A^{BL}
\end{array}\right).
\end{eqnarray}
Then $\sin^2\theta_{_W}'$ can be written as
\begin{eqnarray}
\sin^2\theta_{_W}'=\frac{1}{2}-\frac{(g_{_{YB}}^2-g_{_1}^2-g_{_2}^2)x^2+
4g_{_B}^2}{2\sqrt{(g_{_{YB}}^2+g_{_1}^2+g_{_2}^2)x^4+8g_{_B}^2(g_{_{YB}}^2-g_{_1}^2-g_{_2}^2x^2)+16g_{_B}^2}},
\end{eqnarray}
where $x=\frac{v}{u}$. Compared with the MSSM, this $Z-Z'$ mixing makes new contributions to the $t\rightarrow cZ$ decay channel, and the order of magnitude of $\sin\theta_{_W}'$ is about $\mathcal{O}(10^{-3})$~\cite{Barate:1999qx,Abreu:2000ap,King:2005jy}. The exact eigenvalues of Eq.(\ref{gauge matrix}) are given by
\begin{eqnarray}
&&\qquad\;\quad\;m_\gamma^2=0,\nonumber\\
&&\qquad\;\quad\;m_{Z,{Z^{'}}}^2=\frac{1}{8}\Big((g_{_1}^2+g_2^2+g_{_{YB}}^2)v^2+4g_{_B}^2u^2 \nonumber\\
&&\qquad\;\qquad\;\qquad\;\mp\sqrt{(g_{_1}^2+g_{_2}^2+g_{_{YB}}^2)^2v^4+8(g_{_{YB}}^2-g_{_1}^2-
g_{_2}^2)g_{_B}^2v^2u^2+16g_{_B}^4u^4}\Big),
\end{eqnarray}
In addition, the charged Higgs boson and $W$ gauge boson mass can be written as
\begin{eqnarray}
&&\qquad\;\quad\;m_{H^\pm}^2=\frac{4B_\mu(1+\tan\beta^2)+g_2^2 v_1^2\tan\beta(1+\tan\beta^2)}{4\tan\beta},\nonumber\\
&&\qquad\;\quad\;m_W^2=\frac{1}{4}g_{_2}^2v^2.
\end{eqnarray}

Then the gauge kinetic mixing leads to the mixing between the $H_1^1,\;H_2^2,\;\tilde{\eta_1},\;\tilde{\eta_2}$ at the tree level. In the basis (${\rm Re}H_1^1$, ${\rm Re}H_2^2$, ${\rm Re}\tilde{\eta}_1$, ${\rm Re}\tilde{\eta}_2$),
the tree level mass squared matrix for scalar Higgs bosons is given by
\begin{eqnarray}
&&M_h^2=u^2\times\nonumber\\
&&\left(\begin{array}{*{20}{c}}
{\frac{1}{4}\frac{g^2 x^2}{1+\tan\beta^2}+n^2\tan\beta}&{-\frac{1}{4}g^2\frac{x^2\tan\beta}{1+\tan^2\beta}}-n^2&
{\frac{1}{2}g_{_B}g_{_{YB}}\frac{x}{T}}&
{-\frac{1}{2}g_{_B}g_{_{YB}}\frac{x\tan\beta'}{T}}\\ [6pt]
{-\frac{1}{4}g^2\frac{ x^2\tan\beta}{1+\tan^2\beta}}-n^2&{\frac{1}{4}\frac{g^2\tan^2\beta x^2}{1+\tan\beta^2}+\frac{n^2}{\tan\beta}}&
{\frac{1}{2}g_{_B}g_{_{YB}}\frac{x\tan\beta}{T}}&{\frac{1}{2}g_{_B}g_{_{YB}}\frac{x\tan\beta\tan\beta'}{T}}\\ [6pt]
{\frac{1}{2}g_{_B}g_{_{YB}}\frac{x}{T}}&{\frac{1}{2}g_{_B}g_{_{YB}}\frac{x\tan\beta}{T}}&{\frac{g_{_B}^2}{1+\tan^2\beta'}+\tan\beta'N^2}&
{-g_{_B}^2\frac{\tan\beta'}{1+\tan^2\beta'}-N^2}\\ [6pt]
{-\frac{1}{2}g_{_B}g_{_{YB}}\frac{x\tan\beta'}{T}}&{\frac{1}{2}g_{_B}g_{_{YB}}\frac{x\tan\beta\tan\beta'}{T}}&
{-g_{_B}^2\frac{\tan\beta'}{1+\tan^2\beta^{'}}-N^2}&{g_{_B}^2\frac{\tan^2\beta'}{1+\tan^2\beta'}+\frac{N^2}{\tan\beta'}}
\end{array}\right)
\end{eqnarray}
where $g^2=g_{_1}^2+g_{_2}^2+g_{_{YB}}^2$, $T=\sqrt{1+\tan^2\beta}\sqrt{1+\tan^2\beta'}$,
$n^2=\frac{{\rm Re}B\mu}{u^2}$ and $N^2=\frac{{\rm Re}B\mu^{'}}{u^2}$, respectively. Compared the MSSM, this new mixing in the B-LSSM can affect the theoretical prediction of the process $t\rightarrow ch$.

Including the leading-log radiative corrections from stop and top quark, the mass of the SM-like Higgs boson can be written as~\cite{HiggsC1,HiggsC2,HiggsC3}
\begin{eqnarray}
&&\Delta m_h^2=\frac{3m_t^4}{2\pi v^2}\Big[\Big(\tilde{t}+\frac{1}{2}+\tilde{X}_t\Big)+\frac{1}{16\pi^2}\Big(\frac{3m_t^2}{2v^2}-32\pi\alpha_3\Big)\Big(\tilde{t}^2
+\tilde{X}_t \tilde{t}\Big)\Big],\nonumber\\
&&\tilde{t}=log\frac{M_S^2}{m_t^2},\qquad\;\tilde{X}_t=\frac{2\tilde{A}_t^2}{M_S^2}\Big(1-\frac{\tilde{A}_t^2}{12M_S^2}\Big),\label{higgs corrections}
\end{eqnarray}
where $\alpha_3$ is the strong coupling constant, $M_S=\sqrt{m_{\tilde t_1}m_{\tilde t_2}}$ with $m_{\tilde t_{1,2}}$ denoting the stop masses, $\tilde{A}_t=A_t-\mu \cot\beta$ with $A_t=T_{u,33}$ being the trilinear Higgs stop coupling and $\mu$ denoting the Higgsino mass parameter. Then the SM-like Higgs mass can be written as
\begin{eqnarray}
&&m_h=\sqrt{(m_{h_1}^0)^2+\Delta m_h^2},\label{higgs mass}
\end{eqnarray}
where $m_{h_1}^0$ denotes the lightest tree-level Higgs mass.

Meanwhile, additional D-terms contribute to the mass matrices of the squarks and sleptons, and down type squarks affect the subsequent analysis. On the basis $(\tilde d_L, \tilde d_R)$, mass matrix for down type squarks is given by
\begin{eqnarray}
&&m_{\tilde d}^2=
\left(\begin{array}{cc}m_{dL},&\frac{1}{\sqrt2}(v_1 T_d^\dagger-v_2\mu Y_d^\dagger)\\\frac{1}{\sqrt2}(v_1 T_d-v_2\mu^* Y_d),&m_{dR}\end{array}\right),
\end{eqnarray}
\begin{eqnarray}
&&m_{dL}=\frac{1}{24}\Big(2g_{_B}(g_{_B}+g_{_{YB}})(u_2^2-u_1^2)+(3g_2^2+g_1^2+g_{_{YB}}^2+
g_{_B}g_{_{YB}})(v_2^2-v_1^2)\Big)\nonumber\\
&&\qquad\;\quad\;+m_{\tilde q}^2+\frac{v_1^2}{2}Y_d^\dagger Y_d,\nonumber\\
&&m_{dR}=\frac{1}{24}\Big(2g_{_B}(g_{_B}-2g_{_{YB}})(u_1^2-u_2^2)+2(g_1^2+g_{_{YB}}^2-
\frac{1}{2}g_{_B}g_{_{YB}})(v_2^2-v_1^2)\Big)\nonumber\\
&&\qquad\;\quad\;+m_{\tilde d}^2+\frac{v_1^2}{2}Y_d^\dagger Y_d.
\end{eqnarray}
It can be noted that new gauge coupling constants $g_{_B}$ and $g_{_{YB}}$, with respect to the MSSM, affect the masses of down type squarks significantly when $u$ is large.
%%%%%%%%%%%%%%%%%%%%%%%%%%%%%END REVISED%%%%%%%%%%%%%%%%%%%%%%%%%%%%%%%%%%%%%

\section{Theoretical calculation on $t\rightarrow c\gamma, cg, cZ$ and $ch$ processes\label{sec3}}
\begin{figure}
\setlength{\unitlength}{1mm}
\centering
\includegraphics[width=5.5in]{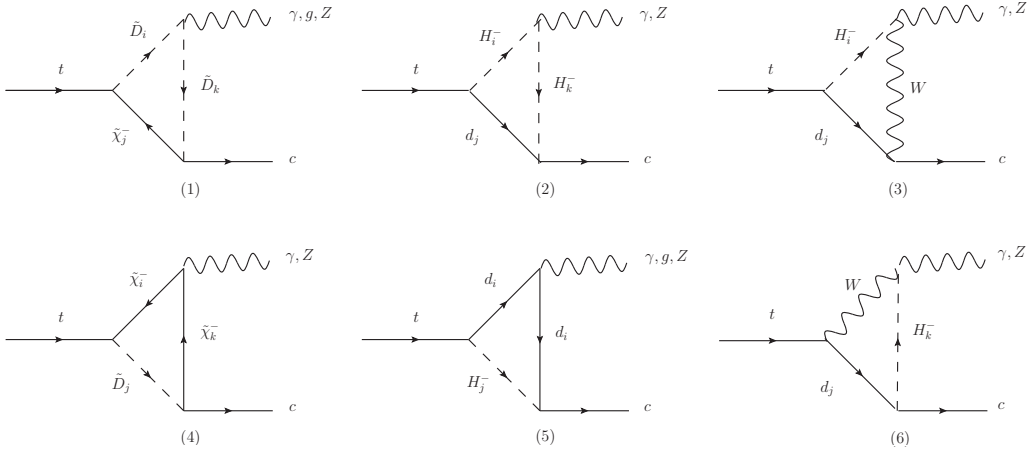}
\vspace{0cm}
\caption[]{The dominating Feynman diagrams contributing to $t\rightarrow c\gamma,\;g,\;Z$ in the B-LSSM} \label{fig1}
\end{figure}

\begin{figure}
\setlength{\unitlength}{1mm}
\centering
\includegraphics[width=5.5in]{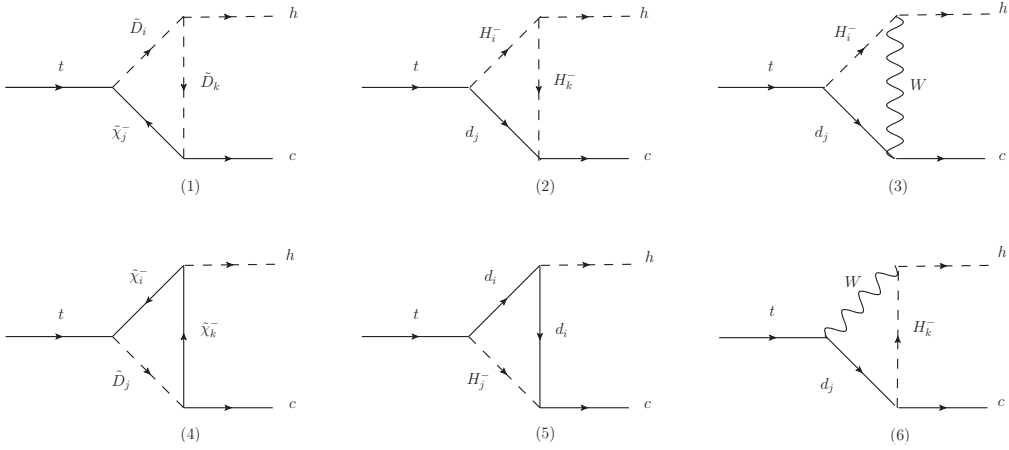}
\vspace{0cm}
\caption[]{The Feynman diagrams contributing to $t\rightarrow ch$ in the B-LSSM} \label{fig4}
\end{figure}

\indent\indent
In this section, we analyze one-loop radiative corrections to the rare decay processes
of top quark $t\rightarrow c\gamma, cg, cZ$ and $ch$ in the B-LSSM. Here the contributions
from self-energy are attributed to the renormalization of external quarks wave functions.
The dominating triangle diagrams contributing to the rare top quark decay processes $t\rightarrow c\gamma, cg, cZ, ch$
are presented in Figs.~\ref{fig1}-\ref{fig4}, respectively.

In the B-LSSM, the corresponding amplitude for the rare decay process $t\rightarrow c\gamma,\;g,\;Z$ is written as
\begin{eqnarray}
&&\mathcal{M}_{tcV}=(4\pi)^{-2}\varepsilon^{\mu} \bar{u}_c(p^{'}){\Big(}\kappa_{V L1}\gamma_\mu P_L
+i\kappa_{V L2}\sigma_{\mu\nu}q^\nu P_L+(L\rightarrow R){\Big)}u_t(p),
\label{tcgamma1}
\end{eqnarray}
where $V$ denotes $\gamma,\;g,\;Z$, $u_t$ and $u_c$ denote the wave functions of top quark and charm quark,
$p$ is the momentum of top quark, $p^{'}$ is the momentum of charm quark,
$q$ is the momentum of vector boson, and $\varepsilon^\mu$ denotes the polarization vector of
photon, gluon and Z boson. Correspondingly, the coefficients $\kappa_{V Li}$ and $\kappa_{V Ri}\;(i=1,2)$
are originating from those Feynman diagrams in Fig.\ref{fig1}. The picture shows that, compared the MSSM, the new definition of the down type squark mass matrix and the corresponding rotation matrix affect the predictions on the processes $t\rightarrow c\gamma, cg, cZ$ in the B-LSSM. In addition, the $Z-Z'$ mixing in Eq.(\ref{gauge matrix}) also makes new contributions to the $t\rightarrow cZ$ decay channel.

Then in order to explain how the calculation of the feynman diagrams in Fig.~\ref{fig1} has been performed, we will take the calculation of Fig.~\ref{fig1}(1) below for example. The corresponding amplitude can be written as
\begin{eqnarray}
&&\mathcal{M}_{tcV}^{(1)}=\varepsilon^{\mu} \bar{u}_c(p^{'})\mu^{4-D}\int \frac{d^Dk}{(2\pi)^D}(iA_{\bar c\tilde{D}_k{\tilde\chi^+_j}L}P_L+iA_{\bar c\tilde{D}_k{\tilde\chi^+_j}R}P_R)\frac{i}{p\!\!\!/-k\!\!\!/-m_{\tilde\chi^+_j}}(iA_{\bar {\tilde\chi^+_j}\tilde{D}_itL}P_L\nonumber\\
&&\qquad\;\quad\;+iA_{\bar {\tilde\chi^+_j}\tilde{D}_itR}P_R)\frac{i}{k^2-m_{\tilde{D}_i}^2}\Big(iB_{V\tilde{D}_i\tilde{D}_k}(-2k_\mu+q_\mu)\Big)\frac{i}{(k-q^2)-m_{\tilde{D}_k}^2}u_t(p),
\label{tcq1}
\end{eqnarray}
where $A_{\bar c\tilde{D}_k{\tilde\chi^+_j}L,R},\;A_{\bar {\tilde\chi^+_j}\tilde{D}_itL,R},\;B_{V\tilde{D}_i\tilde{D}_k}$ denote the constant parts of the interaction vertex about $\bar c\tilde{D}_k{\tilde\chi^+_j},\;\bar {\tilde\chi^+_j}\tilde{D}_it,\;V\tilde{D}_i\tilde{D}_k$ respectively, $L$ and $R$ in subscript denote the left-hand part and right-hand part, and all of them can be got through SARAH. Apply the transverse wave condition and Dirac equation, the amplitude can be simplify as
\begin{eqnarray}
&&\mathcal{M}_{tcV}^{(1)}=\frac{1}{16\pi^2}\varepsilon^{\mu} \bar{u}_c(p^{'})\Big(\kappa_{V L1}^{(1)}\gamma_\mu P_L
+i\kappa_{V L2}^{(1)}\sigma_{\mu\nu}q^\nu P_L+(L\rightarrow R)\Big)u_t(p)
\label{tcq2}
\end{eqnarray}
where
\begin{eqnarray}
&&\kappa_{V L1}^{(1)}=-i B_{V\tilde{D}_i\tilde{D}_k}\Big[\Big(m_tm_c(-C_2-C_{22})-2C_{00}\Big)A_{\bar {\tilde\chi^+_j}\tilde{D}_itR}A_{\bar c\tilde{D}_k{\tilde\chi^+_j}L}+\Big(m_t^2(-C_2-C_{22}\nonumber\\
&&\qquad\;\quad\;-C_{12})+m_c^2C_{12}\Big)A_{\bar {\tilde\chi^+_j}\tilde{D}_itL}A_{\bar c\tilde{D}_k{\tilde\chi^+_j}R}-m_{\tilde\chi^+_j}m_cC_2A_{\bar {\tilde\chi^+_j}\tilde{D}_itL}A_{\bar c\tilde{D}_k{\tilde\chi^+_j}L}-\nonumber\\
&&\qquad\;\quad\;m_{\tilde\chi^+_j}m_tC_2A_{\bar {\tilde\chi^+_j}\tilde{D}_itR}A_{\bar c\tilde{D}_k{\tilde\chi^+_j}R}\Big],\\
&&\kappa_{V L2}^{(1)}=-i B_{V\tilde{D}_i\tilde{D}_k}\Big[m_t(-C_2-C_{22}-C_{12})A_{\bar {\tilde\chi^+_j}\tilde{D}_itR}A_{\bar c\tilde{D}_k{\tilde\chi^+_j}L}-m_{\tilde\chi^+_j}C_2A_{\bar {\tilde\chi^+_j}\tilde{D}_itL}A_{\bar c\tilde{D}_k{\tilde\chi^+_j}L}\nonumber\\
&&\qquad\;\quad\;+m_cC_{12}A_{t\tilde{D}_i{\tilde\chi^+_j}L}A_{c\tilde{D}_k{\tilde\chi^+_j}R}\Big],\\
&&\kappa_{V R1,2}^{(1)}=\kappa_{V L1,2}^{(1)} (L\leftrightarrow R).
\end{eqnarray}
Here $C_0, C_1, C_2, C_{00}, C_{11}, C_{12}, C_{22}$ are Passarino-Veltman scalar functions~\cite{Denner}, and the argument above is $[m_V^2,m_c^2,m_t^2,m_{\tilde{D}_i}^2,m_{\tilde{D}_k}^2,m_{\tilde\chi^+_j}^2]$. The other diagrams corresponding to $t\rightarrow cV$ can be calculated similarly and contribute to the operators $\gamma_\mu P_{L,R}$ and $i\sigma_{\mu\nu}q^\nu P_{L,R}$.

In the effective coupling for $h\bar{c}t$, there are only two effective operators:
\begin{eqnarray}
&&\mathcal{M}_{tch}=(4\pi)^{-2}\bar{u}_c(p^{'})(\kappa_{hL}P_L+\kappa_{hR}P_R)u_t(p)\;,
\label{Mtch}
\end{eqnarray}
the coefficients $\kappa_{hL},\;\kappa_{hR}$ are originating from those Feynman diagrams in Fig.~\ref{fig4}:
\begin{eqnarray}
&&\kappa_{hL}=\sum\limits_{a=1}^{6}\kappa_{hL}^{(a)},
\nonumber\\
&&\kappa_{hR}=\sum\limits_{a=1}^{6}\kappa_{hR}^{(a)},
\end{eqnarray}
where the contributions $\kappa_{hL}^{(a)},\;\kappa_{hR}^{(a)}$ depend on the relevant Feynman diagrams in the model. Fig.~\ref{fig4} shows that, except for the new contributions from down type squarks, the mixing between the Higgs doublets and the exotic singlets $\tilde{\eta}_{1,2}$ also affects the $t\rightarrow ch$ decay channel. And we will take the calculation of Fig.~\ref{fig4}(1) below for example. The amplitude can be given as
\begin{eqnarray}
&&\mathcal{M}_{tch}^{(1)}=\bar{u}_c(p')\mu^{4-D}\int \frac{d^Dk}{(2\pi)^D}(iA_{\bar c\tilde{D}_k{\tilde\chi^+_j}L}P_L+iA_{\bar c\tilde{D}_k{\tilde\chi^+_j}R}P_R)\frac{i}{p\!\!\!/-k\!\!\!/-m_{\tilde\chi^+_j}}(iA_{\bar {\tilde\chi^+_j}\tilde{D}_itL}P_L+\nonumber\\
&&\qquad\;\quad\;iA_{\bar {\tilde\chi^+_j}\tilde{D}_itR}P_R)\frac{i}{k^2-m_{\tilde{D}_i}^2}(iB_{h\tilde{D}_i\tilde{D}_k})\frac{i}{(k-q^2)-m_{\tilde{D}_k}^2}u_t(p),
\label{tch1}
\end{eqnarray}
Apply the Dirac equation, the amplitude can be simplify as
\begin{eqnarray}
&&\mathcal{M}_{tch}^{(1)}=\frac{1}{16\pi^2}\bar{u}_c(p')\Big(\kappa_{hL}P_L+\kappa_{hR}P_R\Big)u_t(p)£¬
\label{tch2}
\end{eqnarray}
where
\begin{eqnarray}
&&\kappa_{hL}^{(1)}=-i B_{h\tilde{D}_i\tilde{D}_k}\Big[m_t(C_0+C_1+C_2)A_{\bar f\tilde{D}_itR}A_{\bar c\tilde{D}_k{\tilde\chi^+_j}L}-m_cC_1A_{\bar {\tilde\chi^+_j}\tilde{D}_itL}A_{\bar c\tilde{D}_k{\tilde\chi^+_j}R}\\
&&\qquad\;\quad\;+m_{\tilde\chi^+_j}C_0A_{\bar {\tilde\chi^+_j}\tilde{D}_itL}A_{\bar c\tilde{D}_k{\tilde\chi^+_j}L}\Big],\\
&&\kappa_{hR}^{(1)}=\kappa_{hL}^{(1)} (L\leftrightarrow R).
\end{eqnarray}
the argument in Passarino-Veltman scalar functions above is $[m_h^2,m_c^2,m_t^2,m_{\tilde{D}_i}^2,m_{\tilde{D}_k}^2,m_{\tilde\chi^+_j}^2]$. The other diagrams corresponding to $t\rightarrow ch$ can be calculated similarly and contribute to the operators $P_L$ and $P_R$.

Based on Eq.(\ref{tcgamma1}) and Eq.(\ref{Mtch}), the corresponding branching ratios of the rare decay processes of top quark respectively read as
\begin{eqnarray}
&&Br(t\rightarrow cV)=\frac{|\mathcal{M}_{tcV}|^2\sqrt{((m_t+m_V)^2-m_c^2)((m_t-m_V)^2-m_c^2)}}{32\pi m_t^3\Gamma_{total}},
\nonumber\\
&&Br(t\rightarrow ch)=\frac{|\mathcal{M}_{tch}|^2\sqrt{((m_t+m_h)^2-m_c^2)((m_t-m_h)^2-m_c^2)}}{32\pi m_t^3\Gamma_{total}}.
\end{eqnarray}
where $\Gamma_{total}=1.4$GeV~\cite{17} is the total decay width of top quark.

\section{Numerical analyses\label{sec4}}
\indent\indent
In this section, we present the numerical results of $Br(t\rightarrow c\gamma,\;cg,\;cZ,\;ch)$
with the help of LoopTools and FeynCalc\cite{FC1,FC2}. The relevant SM input parameters are
chosen as $m_W=80.385{\rm GeV},\;m_Z=90.19{\rm GeV},\;\alpha_{em}(m_Z)=1/128.9,\;\alpha_s(m_Z)=0.118,
\;m_t=173.5{\rm GeV},\;m_c=1.275{\rm GeV}$. Meanwhile the CKM matrix is~\cite{17}
\begin{eqnarray}
&&\left(\begin{array}{*{20}{c}}
{0.97417}&{0.2248}&{4.09\times10^{-3}}\\ [6pt]
{-0.22}&{0.995}&{4.05\times10^{-2}}\\ [6pt]
{8.2\times10^{-3}}&{-4\times10^{-2}}&{1.009}
\end{array}\right).
\end{eqnarray}

The updated experimental data~\cite{newZ} on searching $Z^\prime$ indicates
$M_{Z^{'}}\geq4.05{\rm TeV}$ at 95\% Confidence Level (CL),
and Refs.~\cite{20,21} give us an upper bound on the ratio between
the $Z^{'}$ mass and its gauge coupling at 99\% CL as
\begin{eqnarray}
&&M_{Z^{'}}/g_{_B}\geq6{\rm TeV}\;.
\end{eqnarray}
In order to coincide with the experimental data, we choose $M_{Z^{'}}=4.2{\rm TeV}$ in our numerical analysis, then the scope of $g_{_B}$ is limited to $0<g_{_B}\leq0.7$. The LHC experimental data also constrain $\tan\beta^{'}<1.5$. Considering the constraints from the experiments~\cite{limit1}, for those parameters in Higgsino and gaugino sectors, we appropriately fix $M_1=500{\rm GeV},\;M_2=600{\rm GeV},\;M_{BB^{'}}=500{\rm GeV},\;M_{BL}=600{\rm GeV},\;\mu=700{\rm GeV},\;\mu^{'}=800{\rm GeV}$. For simplify, we set $m_{H^\pm}=2{\rm TeV},\;B_\mu'=5\times10^5{\rm GeV}^2,\;T_u=T_d=diag(1, 1, A_t) \rm TeV$. In addition, the first two generations of squarks are strongly constrained by direct searches at the LHC\cite{ATLAS.PRD,CMS.JHEP} and the third generation squark masses are not constrained by the LHC as strong as the first two generations. Therefore we take $m_{\tilde{q}}^2=m_{\tilde{d}}^2=m_{\tilde{u}}^2=diag(2{\rm TeV},2{\rm TeV},m_{\tilde b})$, and the discussion about the observed Higgs signal in Ref.\cite{Basso:2012tr,C.S} limits $m_{\tilde b}\gtrsim1.5{\rm TeV}$.

It's well known that the experimental observation on $Br(\bar B\rightarrow X_S \gamma)$ limits the relevant parameters strongly, hence we consider the constraint from $\bar B\rightarrow X_s\gamma$ in this work. In addition, we further consider the constraint from $B_s^0\rightarrow\mu^+\mu^-$, which might also limit our numerical analyses\cite{Yang:2018fvw}. The latest experimental data for $Br(\bar B\rightarrow X_S \gamma)$ and $Br(B_s^0\rightarrow \mu^+\mu^-)$ read~\cite{17}
\begin{eqnarray}
&&Br(\bar B\rightarrow X_S \gamma)=(3.49\pm0.19)\times 10^{-4},\\
&&Br(B_s^0\rightarrow \mu^+\mu^-)=(2.9_{-0.6}^{+0.7})\times10^{-9}.
\end{eqnarray}

\begin{figure}
\setlength{\unitlength}{1mm}
\centering
\includegraphics[width=3.1in]{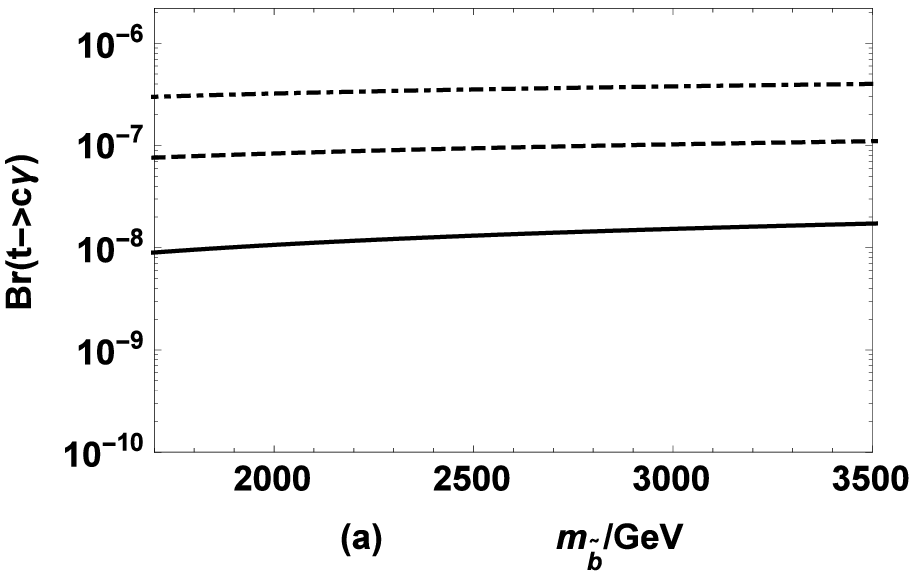}%
\vspace{0.5cm}
\includegraphics[width=3.1in]{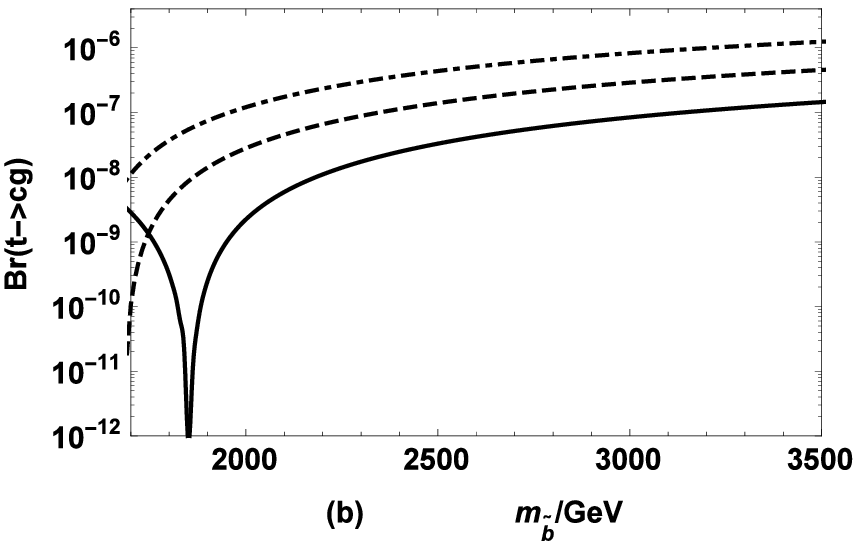}
\vspace{0cm}
\par
\hspace{-0.in}
\includegraphics[width=3.1in]{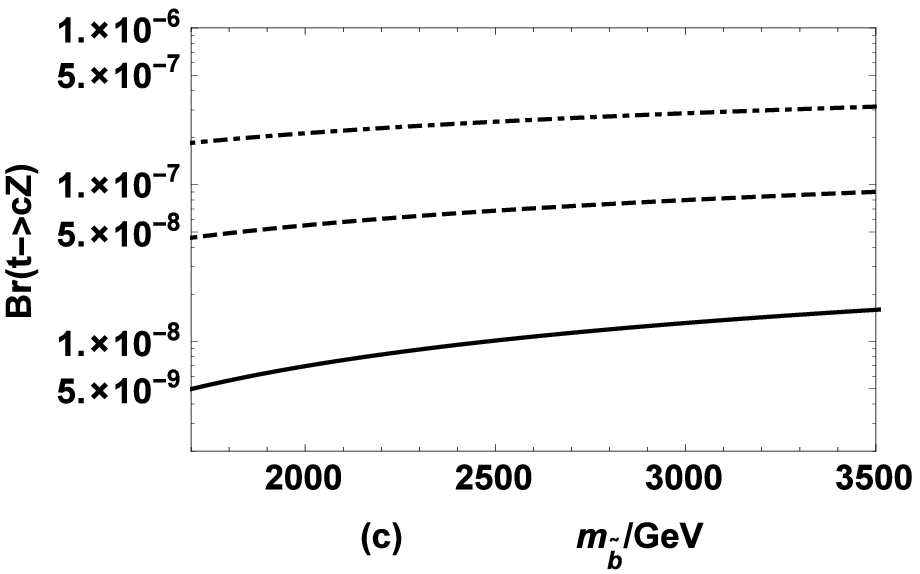}%
\vspace{0.5cm}
\includegraphics[width=3.1in]{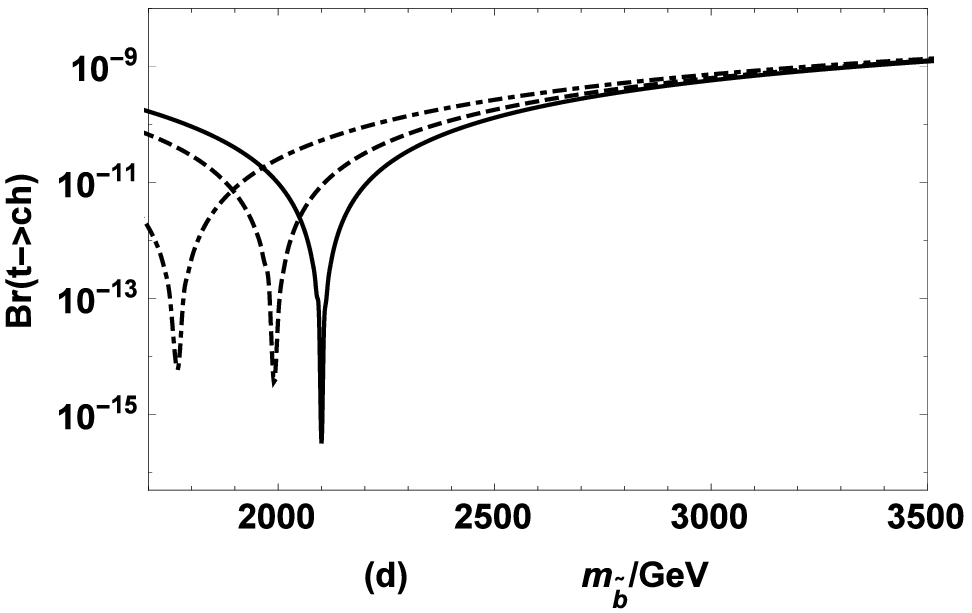}
\vspace{0cm}
\caption[]{$Br(t\rightarrow c\gamma)$(a), $Br(t\rightarrow cg)$(b), $Br(t\rightarrow cZ)$(c), $Br(t\rightarrow ch)$(d) versus $m_{\tilde b}$ for $\tan\beta=15\;({\rm solid\;line}),\;\tan\beta=25\;({\rm dashed\;line}),\;\tan\beta=35$ ({\rm dot-dashed line}) are plotted.}
\label{figm0}
\end{figure}

\begin{figure}
\setlength{\unitlength}{1mm}
\centering
\includegraphics[width=3.1in]{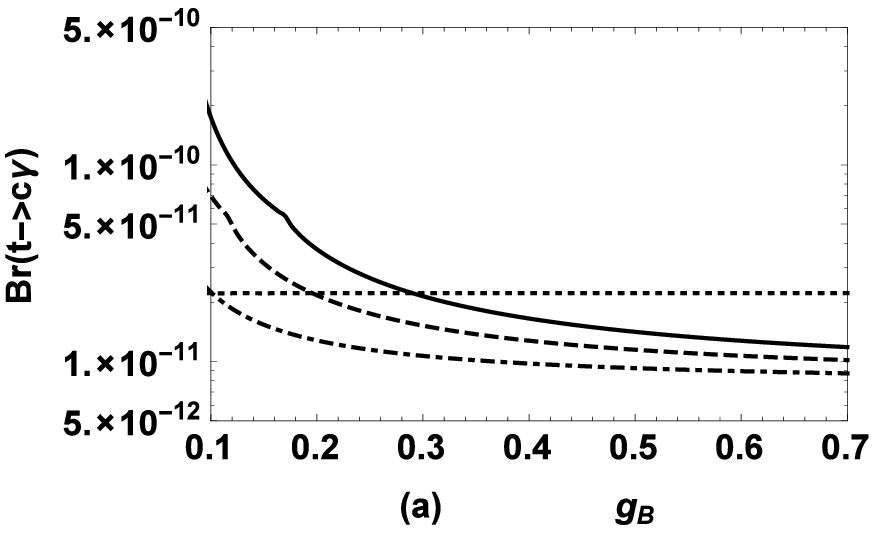}%
\vspace{0.5cm}
\includegraphics[width=3.1in]{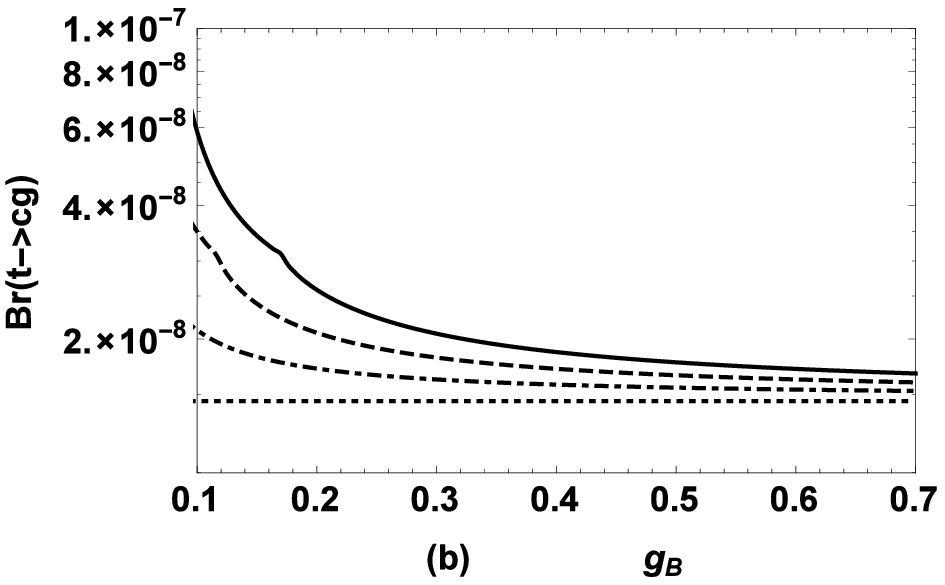}
\vspace{0cm}
\par
\hspace{-0.in}
\includegraphics[width=3.1in]{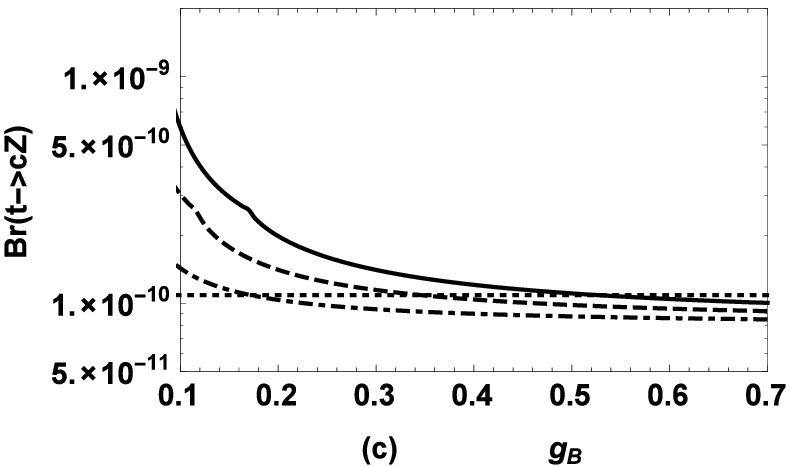}%
\vspace{0.5cm}
\includegraphics[width=3.1in]{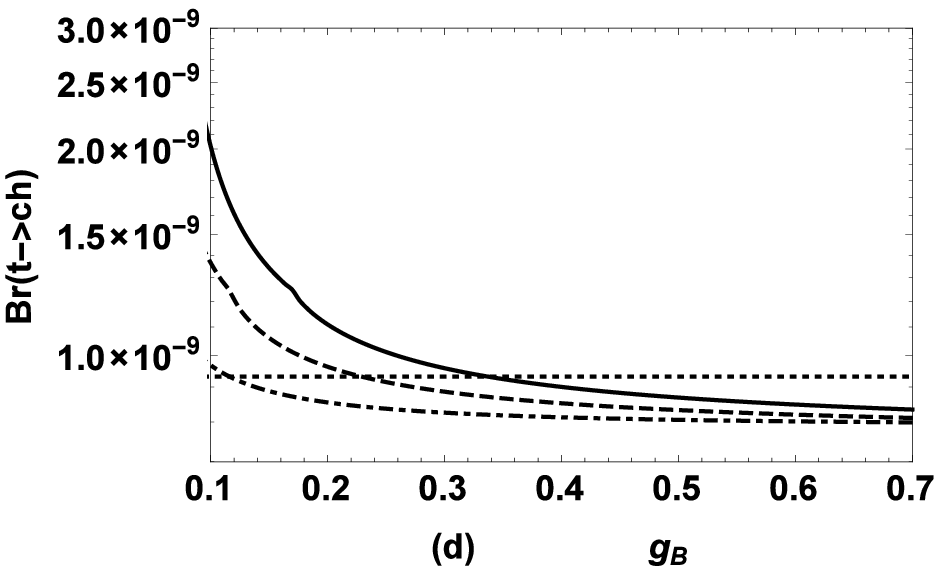}
\vspace{0cm}
\caption[]{$Br(t\rightarrow c\gamma)$(a), $Br(t\rightarrow cg)$(b), $Br(t\rightarrow cZ)$(c), $Br(t\rightarrow ch)$(d) versus $g_{_B}$ for $g_{_{YB}}=-0.6\;({\rm solid \;line}),\;g_{_{YB}}=-0.4\;({\rm dashed \;line}),\;g_{_{YB}}=-0.2$ (dot-dashed line) are plotted. The dotted line denotes the MSSM predictions in the same parameter space.}
\label{figgB}
\end{figure}

We also need to consider the constraint of SM-like Higgs boson mass~\cite{limit2}. Taking $\tan\beta'=1.1,\;g_{_B}=0.2,\;g_{_{YB}}=-0.6,\;A_t=-1.5$ and considering the restrictions from B physics and concrete Higgs boson mass, then letting $m_{\tilde b}$ runs from $1.5{\rm TeV}$ to $4{\rm TeV}$ and $\tan\beta$ runs from $2$ to $40$, the allowed region of them are
\begin{eqnarray}
&&10<\tan\beta<40,\;\;\;\;1700{\rm GeV}<m_{\tilde b}<3500{\rm GeV}.
\end{eqnarray}

Then we plot $Br(t\rightarrow c\gamma),\;Br(t\rightarrow cg),\;Br(t\rightarrow cZ)$ and $Br(t\rightarrow ch)$ versus $m_{\tilde b}$ in Fig.~\ref{figm0}(a-d), where the solid line, dashed line, dot-dashed line denote $\tan\beta=15,\;25,\;35$ respectively. From the picture, we can see that $Br(t\rightarrow c\gamma),\;Br(t\rightarrow cZ)$ increase with the increasing of $m_{\tilde b}$ slowly, but $\tan\beta$ affects both of them obviously. And with the increasing of $\tan\beta$, they can reach $5\times10^{-7}$, $4\times10^{-7}$ respectively. However, Fig.~\ref{figm0}(b) shows that $Br(t\rightarrow cg)$ has a sharp decrease when $\tan\beta=15$, and the turning point is around $1850{\rm GeV}$. Due to the fact that the contributions from down type squarks to the branching ratio is cancelled by the contributions from charge Higgs boson at the turning point. In addition, from the mass matrix of down type squarks, we can see that the masses of down type squarks increase with the increasing of $\tan\beta$ or $m_{\tilde b}$, hence the turning point of $m_{\tilde b}$ decreases with the increasing of $\tan\beta$, which results in the turning point less than $1700$GeV when $\tan\beta=25,\;35$. The moving of turning point can be seen directly in Fig.~\ref{figm0}(d). The picture shows that the turning point decreases with the increasing of $\tan\beta$. In addition, with the increasing of $m_{\tilde b}$, the effect of $\tan\beta$ is more negligible to $Br(t\rightarrow ch)$. Since the main contribution to these processes come from down type squarks, $m_{\tilde b}$ affects the numerical results mainly through influencing the masses of the third generation down type squarks. Meanwhile, $\tan\beta$ not only presents in the diagonal sector of the mass matrix, but also dominates the off-diagonal sector, which indicates that $\tan\beta$ affects the numerical results mainly through influencing the mass of the down type squarks and the corresponding rotation matrix in the couplings involve down type squarks.

In order to see how new coupling constants $g_B$ and $g_{YB}$ in the B-LSSM affect $Br(t\rightarrow c\gamma)$, $Br(t\rightarrow cg)$, $Br(t\rightarrow cZ)$ and $Br(t\rightarrow ch)$, we continue to fix $\tan\beta=4,\;\tan\beta'=1.2,\;m_{\tilde b}=1.5{\rm TeV},\;A_t=-2$. Considering the limits from B physics and concrete Higgs mass, the allowed region of $g_{_B}$ and $g_{_{YB}}$ are
\begin{eqnarray}
&&-0.7<g_{_{YB}}<0,\;\;\;\;0.1<g_{_B}<0.7.
\end{eqnarray}
Then we present $Br(t\rightarrow c\gamma),\;Br(t\rightarrow cg),\;Br(t\rightarrow cZ)$ and $Br(t\rightarrow ch)$ varying with $g_{_B}$ in Fig.~\ref{figgB}(a), (b), (c), (d) respectively, where the three lines denote $g_{_{YB}}=-0.6$ (dotdashed line), $g_{_{YB}}=-0.4$ (dashed line) and $g_{_{YB}}=-0.2$ (dot-dashed line). In order to compare with the MSSM, we also plot the MSSM predictions in the same parameter space (dotted line). It can be noted that $Br(t\rightarrow cg)$ in the B-LSSM can exceed the MSSM prediction easily in our chosen parameter space. And $Br(t\rightarrow c\gamma),\;Br(t\rightarrow cZ),\;Br(t\rightarrow ch)$ in the B-LSSM can exceed the MSSM predictions when $g_{_B}$ is small and $|g_{_{YB}}|$ is large. In addition, as $g_{_{YB}}$ approach to zero, all of the branching ratios depend on $g_{_B}$ negligibly, which indicates that the effect of $g_{_B}$ to these four processes is influenced by the strength of gauge kinetic mixing strongly. $g_{_B}$ and $g_{_{YB}}$ affect the numerical results mainly in three ways. Firstly, $g_{_B}$ and $g_{_{YB}}$ affect $Br(t\rightarrow c\gamma,\;g,\;Z,\;h)$ by influencing the down type squark masses and the corresponding rotation matrix, which appears in the couplings involve the down type squarks. Secondly, they make new contributions to $Br(t\rightarrow cZ)$ by the $Z-Z'$ mixing. Thirdly, they affect the theoretical prediction on $Br(t\rightarrow ch)$ by mixing the Higgs doublets with the exotic singlets.

\section{Summary\label{sec5}}
\indent\indent
In the $U(1)_{B-L}$ extension of MSSM, under a minimal flavor violating assumption for the soft breaking terms, we focused on the top quark rare decay processes $t\rightarrow c\gamma, cg, cZ, ch$. Compared with the MSSM, new definition of the down type squark masses can affect the theoretical evaluation on these processes. In addition, the mixing in the scalar sector and $Z$-$Z'$ sector can also make new contributions to $t\rightarrow ch$ and $t\rightarrow cZ$ decay channel respectively. In our used parameter space, the numerical results show that all of these processes are well below the experiment limits. And $\tan\beta$ is a major parameter to the processes $t\rightarrow c\gamma, cg, cZ, ch$, the corresponding branching ratios can be $5\times10^{-7}$, $2\times10^{-6}$, $4\times10^{-7}$, $3\times10^{-9}$ respectively. Simultaneously, new gauge coupling constants $g_{_B},\;g_{_{YB}}$ in the B-LSSM can also affect the numerical results of $Br(t\rightarrow c\gamma,\;cg,\;cZ,\;ch)$.

\begin{acknowledgments}
\indent\indent
The work has been supported by the National Natural
Science Foundation of China (NNSFC) with Grants
No. 11535002, No. 11647120, and No. 11705045,
Natural Science Foundation of Hebei province with
Grants No. A2016201010 and No. A2016201069,
Foundation of Department of Education of Liaoning
province with Grant No. 2016TSPY10, Youth
Foundation of the University of Science and Technology
Liaoning with Grant No. 2016QN11, Hebei Key Lab of
Optic-Eletronic Information and Materials, and the
Midwest Universities Comprehensive Strength Promotion
project.
\end{acknowledgments}

\end{document}